\documentclass{article}
\usepackage{geometry}
\usepackage{natbib}
\usepackage[utf8]{inputenc}
\usepackage[T1]{fontenc}
\usepackage{hyperref}
\usepackage{url}
\usepackage{booktabs}
\usepackage{amsfonts}
\usepackage{amsmath}
\usepackage{amssymb}
\usepackage{xcolor}
\usepackage{graphicx}
\usepackage{float}   % provides [H] placement
\title{Keep the Proof State Live: Snapshotting for Efficient Tactic Search in Lean~4}

\author{%
  Austin Shen \\
  University of Michigan \\
  \texttt{austinshen@umich.edu}
  \and
  Yunong Shi \\
  Amazon Web Services \\
  University of Michigan \\
  \texttt{shiyunon@amazon.com} \\
  \texttt{yunong@umich.edu}
}

\begin{document}
\maketitle

\begin{abstract}
Automated theorem proving systems built on Lean~4 increasingly rely on parallel
tactic search over partially specified proofs, such as those generated by
Draft--Sketch--Prove (DSP) pipelines.
In current systems, each search branch reconstructs a proof state by re-running
elaboration (Lean's process of type-checking and resolving implicit arguments),
leading to substantial per-branch overhead.
In Lean~4 with Mathlib, this cost has two components:
(1)~\emph{import loading}, which deserializes pre-compiled libraries
($\approx$60~s per branch); and
(2)~\emph{theorem-body elaboration}, which re-checks the theorem context up to the
target goal (estimated 18--735~s depending on proof complexity).
Together, these account for $>$99\% of per-branch wall time in a standard
single-machine setting, making portfolio-based search (trying a fixed set of
candidate tactics in parallel across each proof hole) impractical at scale.

We observe that this overhead arises from a mismatch between the structure of proof
search and its execution model: branching is implemented via repeated reconstruction
of proof states rather than direct reuse.
To address this, we introduce \emph{proof-state snapshotting}, which captures the
elaborated proof state once and reuses it across branches.
We implement this mechanism via a small extension to the Lean~4 language server,
enabling lightweight forking of proof states for parallel tactic execution.

Across 48 miniF2F-v2 problems (45 hand-crafted prove-phase benchmarks and 3 full
end-to-end runs), our approach achieves a \textbf{5.6--50$\times$}
wall-time speedup over the standard fallback (average \textbf{14$\times$}, median
\textbf{9.7$\times$} across the 45 hand-crafted benchmarks).
Speedup increases with the number of proof branches.
Tactic execution itself accounts for only a few ms to 500~ms per branch in typical cases
(95th percentile 289~ms; \texttt{aesop} reaches several seconds on hard goals);
the remaining $>$99\% of runtime is elaboration overhead that our approach eliminates.

Our method is orthogonal to import-level caching
(e.g., Kimina Lean Server~\citep{kimina2025}), which avoids repeated import loading
but not theorem-body elaboration.
By eliminating both costs within a theorem, proof-state snapshotting provides a
complementary mechanism for scaling parallel proof search.
The patched Lean binary and the Snapshot-DSP pipeline will be released as open source at~\url{https://github.com/A2DR1/Lean_Snapshot}.
\end{abstract}

%% -----------------------------------------------------------------------
\section{Introduction}
\label{sec:intro}

Automated formal theorem proving with large language models has advanced
rapidly~\citep{leandojo,dsp2023,leancopilot,leanagent,dspplus2025,%
aristotle2025,kimina2025,pantograph,leantree},
with systems achieving 88.9\% on miniF2F~\citep{deepseekproverv2} and
IMO-level performance~\citep{aristotle2025}.
A common pattern in these pipelines is \emph{portfolio-based search}: fix a set
of candidate tactics and try each one in parallel across every proof hole.
The bottleneck is not tactic execution---which takes a few ms to 500~ms in typical cases---but
\emph{branch startup}: the cost of reconstructing proof state from scratch on
every attempt.

In Lean~4 with Mathlib, branch startup has two compounding components.
\emph{Import loading} deserializes pre-compiled \texttt{.olean} libraries
($\approx$60~s, constant across problems).
\emph{Theorem-body elaboration} type-checks the theorem header and all
intermediate steps up to the \texttt{sorry} position (estimated 18--735~s, scaling with
proof complexity).\footnote{%
  The 18~s lower bound is $\min(\text{native\_s}) - T_{\text{import}}
  \approx 78\text{ s} - 60\text{ s}$ (using estimated $T_{\text{import}}\approx 60$~s),
  from the simplest benchmark problem (\texttt{mathd\_algebra\_209}).
  The 735~s upper bound is back-computed from \texttt{mathd\_numbertheory\_328}
  (28 branches, $W\!=\!2$ workers):
  $\lceil 28/2 \rceil \times (T_{\text{import}} + T_{\text{body}}) = 11{,}127~\text{s}
  \Rightarrow T_{\text{body}} \approx 735~\text{s}$.}
Together these consume $>$99\% of per-branch wall time.
A portfolio of 7 tactics applied to 4 \texttt{sorry} holes yields 28
branches---roughly 17.5~minutes of pure overhead on simple theorems (with $W\!=\!2$
concurrent workers; 35~minutes if fully sequential), and hours on complex ones.
At the scale of DSP-style pipelines, which may propose 100 draft candidates per
theorem, this makes portfolio-based search infeasible on a single machine.

The root cause is a mismatch between how the Lean server manages state internally
and how external systems access it.
The Lean server (\texttt{lean --server}) already computes and caches the
elaborated proof state at every source position---this is what powers live
diagnostics in VS~Code and Emacs.
That state is never discarded between edits; it is simply not exposed to external
callers.
Every branch therefore reconstructs state the server already holds.

\begin{figure}[!t]
\centering
\includegraphics[width=\linewidth]{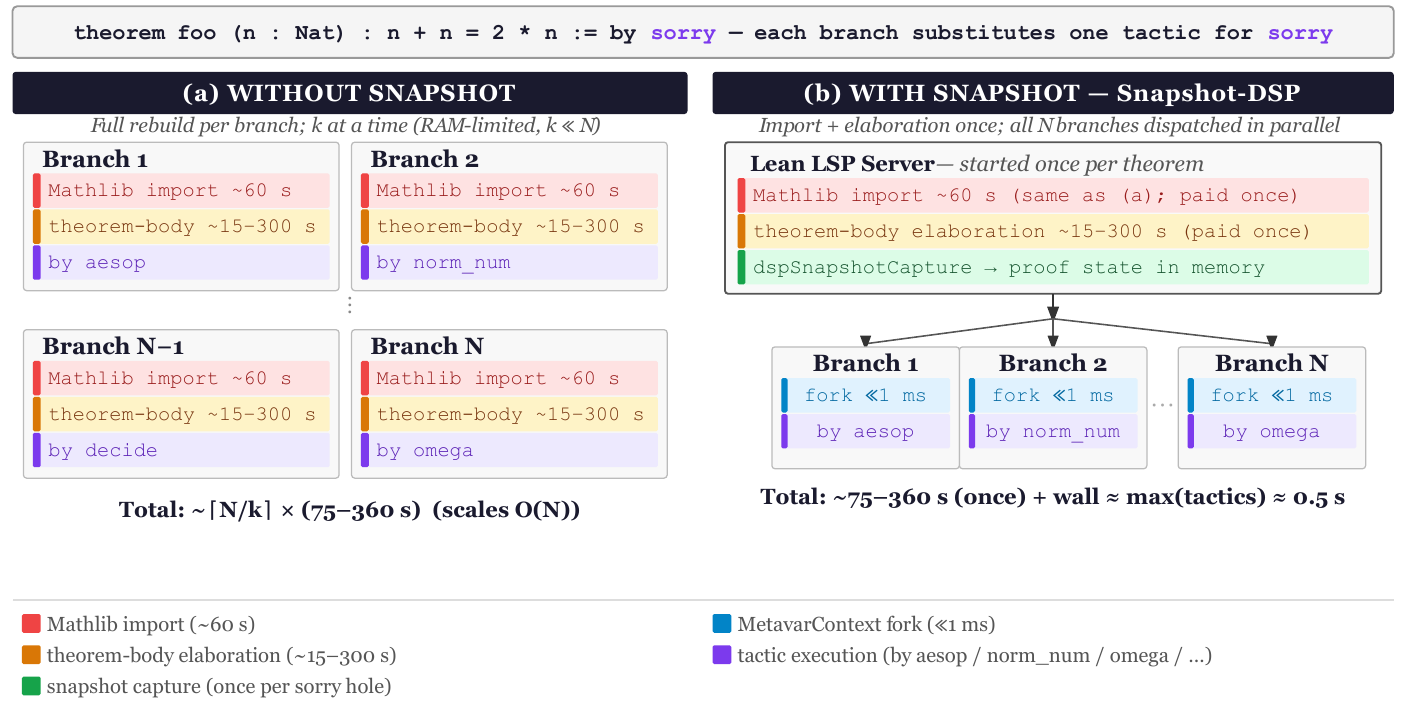}
\caption{%
  \textbf{(a) Fallback (Level~0):} every branch independently reloads Mathlib and
  re-elaborates the theorem body (75--795~s per branch depending on theorem
  complexity; 75~s for simple theorems, up to 795~s for the hardest benchmark),
  yielding $O(N)$ total overhead.
  The diagram shows $W\!=\!2$ concurrent workers; in general $W \approx \lfloor\text{RAM}/3\,\text{GB}\rfloor$,
  so $W \ll N$ on any single machine.
  \textbf{(b) Snapshot-DSP (Level~2):} import loading and elaboration are paid once;
  all $N$ branches are dispatched in parallel and together add only 4--13~s
  in LSP overhead and tactic execution on top of the one-time capture.
  The tactic portfolio \{\texttt{aesop}, \texttt{norm\_num}, \texttt{decide}, \texttt{omega}, \ldots\}
  is a fixed, statically dispatched set---every tactic is tried on every hole
  independently and simultaneously.}
\label{fig:comparison}
\end{figure}

We address this with \emph{proof-state snapshotting}.
We extend \texttt{Lean.Server.FileWorker} with three JSON-RPC methods---%
\texttt{dspSnapshotPing}, \texttt{dspSnapshotCapture}, and
\texttt{dspSnapshotBranch}---that make the server's internal elaboration snapshot
an externally reusable branching substrate.
The \texttt{Environment} (all Mathlib constants and type-class instances,
$\approx$2--4~GB) is immutable and shared across branches by reference; only the
\texttt{MetavarContext} (open proof obligations, $\approx$KB) is copied per
branch.
This reduces per-branch memory from $O(\text{GB})$ to $O(\text{KB})$, removing RAM
as a parallelism bottleneck: the fallback requires $\approx$3~GB per concurrent
branch (limiting a typical laptop to $W\!\approx\!2$ workers), whereas snapshotting
allows all $N$ branches (equivalently $B = H \times C$) to run in parallel
regardless of $N$ or available RAM.
Mathlib loads once per theorem; after capture, all $N$ branches are dispatched in
parallel and the full set adds only 4--13~s of combined LSP overhead and
tactic execution on top of the one-time capture cost.
Figure~\ref{fig:comparison} illustrates the difference.

\paragraph{Positioning.}
Level~0 systems (stock DSP via \texttt{lake build}) fully rebuild per branch.
Level~1 systems (e.g., Kimina~\citep{kimina2025}) cache the post-import
\texttt{Environment} across theorems, eliminating import loading but still paying
theorem-body elaboration (estimated 18--735~s) per branch.
This work contributes \textbf{Level~2}: snapshotting the proof state at the
\texttt{sorry} boundary, after theorem-body elaboration, eliminating both costs.
Levels~1 and~2 are complementary and can be combined in a persistent-server
configuration, which we leave to future work.
Separately, Lean already supports intra-branch backtracking via
\texttt{SavedState} checkpoints, used by Pantograph~\citep{pantograph} and
Aesop~\citep{aesop}; this mechanism operates within a single branch and is
orthogonal to the branch-dispatch cost that snapshotting addresses.

\paragraph{Results.}
Across 48 miniF2F-v2 problems, proof-state snapshotting achieves a
\textbf{5.6--50$\times$} wall-time speedup over the fallback (average
\textbf{14$\times$}, median \textbf{9.7$\times$} across the 45 hand-crafted benchmarks).
Speedup scales monotonically with branch count, confirming that the fixed
elaboration cost amortizes as the portfolio grows.

\paragraph{Contributions.}
\begin{enumerate}
\item \textbf{System.} A proof-state snapshotting system comprising Lean~4 LSP
      extensions for capturing and reusing elaborated proof states
      (\texttt{MetavarContext} / \texttt{LocalContext}) across branches, and a
      Python orchestration layer for dispatching tactic portfolios over
      snapshot-based execution (\S\ref{sec:method}).
\item \textbf{Evaluation.} Speedup measurements across 48 miniF2F-v2 problems,
      with cost decomposition confirming that $>$99\% of per-branch wall time is
      elaboration overhead that snapshotting eliminates (\S\ref{sec:eval}).
\item \textbf{Release.} We will release the Snapshot-DSP pipeline, patched Lean
      binary, and evaluation suite as open source at
      \url{https://github.com/A2DR1/Lean_Snapshot}.
\end{enumerate}

%% -----------------------------------------------------------------------
\section{Related Work}
\label{sec:related}

\paragraph{LLM-guided theorem proving.}
A broad family of systems treat Lean as an external black box, submitting tactic
strings and receiving goal states via LSP or a REPL wrapper.
\citet{dsp2023} introduce the Draft--Sketch--Prove pipeline (originally for
Isabelle/Sledgehammer); \citet{leandojo} build LeanDojo with retrieval-augmented
proving; \citet{legoProver} augment search with a growing lemma library;
\citet{deepseekproverv2} achieves 88.9\% on miniF2F with a 671B MoE model.
\citet{dspplus2025} revisit DSP for reasoning-era models, scaling the sketch
generation and parallel prove phase.
To our knowledge, none of these systems identify or address Level~2 as a design
point; all pay per-branch elaboration overhead in the Prove phase.

\paragraph{Large-scale parallel proof search.}
\citet{aristotle2025} achieve gold-medal performance on the International Mathematical
Olympiad using Monte Carlo Graph Search with learned policies and value functions,
deployed across a massive CPU cluster via a custom REPL service with compressed state
serialization.
This system operates at Level~0: branch expansions are dispatched to independent
stateless backends that reconstruct the full proof state on every request;
the paper's compressed state serialization refers to the per-request payload,
not a cached elaboration artifact.
Our elaboration-level snapshots are complementary: Level~2 forking could provide
Aristotle-style search with pre-elaborated snapshots from the IDE's own server,
eliminating startup cost without changing the search algorithm.

\paragraph{Proof environment caching.}
The closest prior work is the Kimina Lean Server~\citep{kimina2025}, which
maintains an LRU cache keyed on import headers: if a new verification request
shares the same \texttt{import} prefix as a cached session, the pre-loaded Lean
process is reused, reporting a \textbf{1.94$\times$} speedup on NuminaMath-LEAN.
Kimina and Snapshot-DSP are complementary: Kimina caches the post-import state
\emph{between theorems} (Level~1); we fork the post-elaboration state
\emph{between branches of the same theorem} (Level~2).
Even inside a Kimina-style persistent server, every branch still pays
theorem-body elaboration---the cost our snapshot eliminates.
The two mechanisms compose, and our current implementation does not yet
exploit Kimina-style inter-theorem caching, leaving that as future work.
LeanInteract~\citep{leaninteract} takes a complementary approach: a persistent
Python-facing REPL that keeps a single Lean process alive across queries,
enabling low-latency proof attempts by submitting complete proofs directly
to a warm environment.
Unlike Snapshot-DSP, LeanInteract is designed for sequential branch dispatch
rather than parallel fork-and-check; the two designs trade off per-branch latency
against parallelism and single-theorem startup cost.

\paragraph{Proof state management.}
Pantograph~\citep{pantograph} provides a machine-to-machine Lean interface that
exposes proof states as structured data (goal factorization, state deletion) rather
than opaque LSP responses.
This is closer to the internal representation than raw LSP, but Pantograph still
dispatches tactics sequentially against a single state: it does not capture a fixed
proof state and fork it for a parallel tactic portfolio, which is the operation
Level~2 requires.
LeanTree~\citep{leantree} decomposes complex goals into independent sub-branches via
a factorized state representation, reporting a 3.4$\times$ accuracy improvement over
black-box single-branch rollout on miniF2F with Llemma-7B.
LeanTree addresses goal structure (how a single proof state is split into sub-goals),
whereas Snapshot-DSP addresses branch dispatch (how cheaply a fixed proof state is
replicated across a tactic portfolio); the two approaches are orthogonal.
Aesop~\citep{aesop} is a white-box best-first tactic that uses \texttt{SavedState}
checkpoints internally for backtracking within a single branch---the intra-branch
axis described in \S\ref{sec:intro}.
It is one of the seven tactics in our portfolio, and its internal use of
\texttt{SavedState} is orthogonal to and unaffected by the Level~2 snapshot.

\paragraph{Lean~4 parallel elaboration.}
Lean~4.19 (May 2025) adds parallel elaboration of theorem bodies, allowing
the compiler to type-check multiple top-level definitions concurrently during a build
(groundwork was laid across releases 4.16--4.18).
This is designed for batch compilation (\texttt{lake build} across a project), not
for runtime proof-state forking in proof search: it does not allow an external
client to capture or branch an in-progress elaboration.
Our snapshot mechanism addresses this gap at runtime and could be upstreamed.

%% -----------------------------------------------------------------------
\section{Method}
\label{sec:method}

The Prove phase operates as a five-step pipeline.
First, the Python orchestrator pings the Lean server to detect snapshot support;
if unavailable, it falls back to \texttt{lake build} per branch.
Second, it finds all \texttt{sorry} positions in the LLM-generated sketch by
parsing the source file.
Third, for each \texttt{sorry} position it captures a snapshot---a handle to the
elaborated proof state at that point.
Fourth, it branches by sending all tactic configs for a given snapshot in a
single batch RPC call; the Lean server fans them out in parallel using Lean's
work-stealing task scheduler (\texttt{IO.asTask}), so wall time for the branch
step is approximately $\max(\text{branch costs})$ rather than their sum.
Fifth, it collects results: a theorem is proved when every hole is closed by at
least one tactic.
The following subsections describe the two main technical components:
the Lean-side snapshot primitives (\S\ref{sec:snapshots}) and how they integrate
with the surrounding DSP pipeline (\S\ref{sec:integration}).

\subsection{Snapshot Capture and Branch Execution}
\label{sec:snapshots}

\paragraph{Background.}
The Lean server (\texttt{lean --server}) maintains a \texttt{FileWorker} per open
document.
As elaboration proceeds, the worker produces a chain of \texttt{Snapshot} objects,
each recording the elaboration state at a source position.
A snapshot holds three components: the \texttt{Environment} (all loaded constants
and type-class instances), the \texttt{MetavarContext} (open proof obligations),
and the \texttt{LocalContext} (local variable bindings).
In a standard IDE workflow, these snapshots exist already---the server uses them to
avoid re-elaborating unchanged lines when the user edits a file.
Snapshot-DSP repurposes this existing machinery for proof search.

\paragraph{Three new RPCs.}
We extend \texttt{Lean.Server.FileWorker} with three JSON-RPC handlers, registered
at server boot before any user document is opened:

\begin{description}
\item[\texttt{\$/lean/dspSnapshotPing}] Capabilities handshake; returns
      \texttt{\{"ok": true\}} if snapshot support is available, allowing the Python
      orchestrator to select the native path or fall back to \texttt{lake build}.

\item[\texttt{\$/lean/dspSnapshotCapture}] Given a file URI and a
      \texttt{(line, character)} position of a \texttt{sorry} token, blocks until
      elaboration reaches that point, records a snapshot reference, and returns
      a stable \texttt{snapshotId} string.

\item[\texttt{\$/lean/dspSnapshotBranch}] Given a \texttt{snapshotId} and an
      array of branch configs (each specifying a tactic string), spawns one
      \texttt{IO.asTask} per branch using Lean's work-stealing scheduler,
      substitutes each tactic for \texttt{sorry}, runs it against the captured goal
      state, and returns an array of
      \texttt{\{"ok": bool, "error": string|null, "cpuSeconds": float\}}.
      All branches receive an independent copy of the \texttt{MetavarContext};
      the shared \texttt{Environment} (Mathlib, $\approx$2--4~GB) is
      read-only and never copied, so spawning a branch is essentially free.
\end{description}

\paragraph{Why forking is cheap.}
Lean's proof state has two components with very different sizes and mutability.
The \texttt{Environment}---all Mathlib constants, type-class instances, simp sets,
and attribute tables---is immutable and approximately 2--4~GB; it is shared across
all branches by reference and never copied.
The \texttt{MetavarContext}---the open metavariable assignments for the current
proof---is the only mutable component and is typically a few kilobytes.
Forking therefore costs only a \texttt{MetavarContext} copy.
Because branches are isolated by construction, the server runs all $N$ branches
in parallel via \texttt{IO.asTask}.
In practice, for a 7-tactic portfolio on a simple arithmetic theorem, branch-dispatch
wall time is $\approx$0.58~s ($\approx$1.20$\times$ the slowest branch at 0.49~s),
versus a sequential prediction of 1.43~s ($= \sum$\,branch CPU times)---a
\textbf{2.5$\times$} reduction in branch-dispatch latency.

\paragraph{Relationship to SavedState.}
Both Level~2 snapshots and Lean's tactic-level \texttt{SavedState} record the same
underlying proof state (open goals, local variable bindings, and metavariable
assignments).
The difference is when and why: \texttt{SavedState} captures between tactic steps
to support intra-branch backtracking within a single sequential search attempt;
Level~2 captures at a \texttt{sorry} boundary to support parallel dispatch across
many independent branches.
Lean's existing \texttt{SavedState} machinery is therefore orthogonal to and
unaffected by the Level~2 snapshot.

\paragraph{Distribution.}
The patched binary (toolchain: \texttt{leanprover/lean4:v4.25.1}) is built
from a fork of the Lean~4 source tree and distributed via \texttt{elan};
the \texttt{lean-toolchain} file in \texttt{examples/minif2f\_workspace}
points to the patched binary.
Standard Lean installations are unaffected.

\subsection{Integration with the DSP Pipeline}
\label{sec:integration}

\paragraph{Sketch generation.}
The Sketch LLM (GPT-4.1-mini) receives the informal draft and formal theorem
statement and returns a complete Lean~4 module with sub-goals and \texttt{sorry}
holes.
A post-processing step splices the exact original formal statement header into the
LLM output before it is passed to the Lean server, preventing Unicode mangling of
identifiers such as \texttt{h\textsubscript{1}} or \texttt{$\mathbb{R}$} with ASCII
approximations that Lean would reject.

\paragraph{Tactic portfolio.}
The default portfolio contains seven standard Lean~4 tactics:
\texttt{aesop}~\citep{aesop}, \texttt{norm\_num}, \texttt{omega}, \texttt{ring},
\texttt{linarith}, \texttt{decide}, and \texttt{simp}.
Each is dispatched as a direct tactic call via \texttt{dspSnapshotBranch};
the portfolio is easily extended.

\paragraph{Fallback path.}
When \texttt{dspSnapshotPing} does not return \texttt{ok}---because the user's Lean
installation is unpatched---the orchestrator falls back transparently to
\texttt{lake build} per branch, writing each tactic variant to a separate copy of
the project directory.
The fallback produces identical results at $\mathcal{O}(B)$ elaboration cost rather
than $\mathcal{O}(1)$.

%% -----------------------------------------------------------------------
\section{Evaluation}
\label{sec:eval}

\subsection{Setup}
We evaluate on 48 problems from miniF2F-v2~\citep{minif2fv2} spanning
number theory, algebra, and competition mathematics (AMC/AIME/IMO).
For three problems the full DSP pipeline (LLM Draft + LLM Sketch + Prove) is run
end-to-end using GPT-4.1-mini;
for the remaining 45 problems, hand-crafted valid Lean~4 sketches with 1--5
\texttt{sorry} holes are used to isolate the Prove-phase wall-time comparison.
49 hand-crafted sketches were initially constructed; 4 failed \texttt{lake build}
due to elaboration errors unrelated to the tactic portfolio and are excluded,
yielding 45 valid Prove-phase benchmarks.
The hand-crafted sketches are designed so that at least one portfolio tactic closes
each hole; both paths run the same complete branch set, so the wall-time comparison
measures overhead reduction uniformly across all 45 problems regardless of per-problem
proof outcome.
\textbf{These benchmarks measure the speed of the tactic search infrastructure, not
proof discovery.}
End-to-end proof-discovery accuracy is addressed only by the three full DSP runs.
All experiments run on a MacBook Pro (Apple M-series, 8~GB RAM, macOS) with the
\texttt{dsp-patched} Lean binary and Lean toolchain \texttt{leanprover/lean4:v4.25.1}.
Each problem is timed twice---once with native snapshots preferred and once
with fallback (\texttt{lake build} per branch) forced---and wall times are recorded.
Fallback concurrency is bounded by available memory (8~GB); effective throughput
is approximately 2 concurrent lake builds ($W\!\approx\!2$) for the 45 hand-crafted
benchmarks. The three end-to-end DSP fallback runs were executed with $W\!=\!1$
(sequential) to reflect realistic deployment conditions where a single user
runs the full pipeline.
Both native and fallback paths fail identically on the excluded inputs (e.g.,
malformed \texttt{have} steps that Lean rejects before reaching any \texttt{sorry} hole),
so their exclusion does not introduce selection bias.

\subsection{Main Results}

Table~\ref{tab:main} reports results aggregated by \texttt{sorry}-hole count.
A key finding is that speedup scales monotonically with hole count: from
$\sim$7.9$\times$ at 1 hole to $\sim$30$\times$ at 5 holes, confirming that
the fixed elaboration cost is amortized across an increasing number of branches.

\begin{table}[h]
\centering
\caption{Wall-time comparison: native snapshot engine vs.\ fallback (\texttt{lake build} per branch).
  $H$ = \texttt{sorry} holes; $B = H \times 7$ branches.
  Top section: full end-to-end DSP runs (LLM Draft + Sketch + Prove).
  Bottom section: aggregated results over 45 hand-crafted Prove-phase benchmarks,
  grouped by hole count.  Speedup range shows min--max within each group.}
\label{tab:main}
\small
\begin{tabular}{lrrrrc}
\toprule
Problem / Group & $H$ & $B$ & Avg Native (s) & Avg Fallback (s) & \textbf{Speedup} \\
\midrule
\multicolumn{6}{l}{\textit{Full DSP pipeline (GPT-4.1-mini)}} \\
\midrule
\texttt{mathd\_numbertheory\_345} & 5 & 35 &  132.8 & 2{,}641.4 & \textbf{19.9$\times$} \\
\texttt{mathd\_numbertheory\_3}   & 3 & 21 &  116.2 & 1{,}572.4 & \textbf{13.5$\times$} \\
\texttt{mathd\_algebra\_478}      & 4 & 28 &  119.9 & 1{,}579.6 & \textbf{13.2$\times$} \\
\midrule
\multicolumn{6}{l}{\textit{Prove-phase benchmark (45 hand-crafted sketches, grouped by hole count)}} \\
\midrule
1 hole \hfill(5 problems)  & 1 &  7 & 196 & 1{,}537 & \textbf{7.9$\times$} (6.3--9.6$\times$) \\
2 holes \hfill(19 problems) & 2 & 14 & 136 & 1{,}291 & \textbf{9.4$\times$} (5.6--19.9$\times$) \\
3 holes \hfill(11 problems) & 3 & 21 & 131 & 1{,}811 & \textbf{13.8$\times$} (6.5--20.2$\times$) \\
4 holes \hfill(7 problems)  & 4 & 28 & 125 & 3{,}478 & \textbf{24.6$\times$} (13.1--50.0$\times$) \\
5 holes \hfill(3 problems)  & 5 & 35 & 141 & 4{,}436 & \textbf{29.8$\times$} (20.1--45.2$\times$) \\
\midrule
\textbf{Average (all 45)} & \textbf{2.6} & \textbf{18.5} & \textbf{140} & \textbf{1{,}995} & \textbf{14.0$\times$} \\
\bottomrule
\end{tabular}
\end{table}

\subsection{Cost Breakdown}

Table~\ref{tab:breakdown} decomposes the per-branch cost across representative
evaluated problems.
Tactic CPU is measured from the \texttt{cpuSeconds} field returned by the Lean
server inside \texttt{dspSnapshotBranch}; overhead fraction is
$(T_{\mathrm{fallback}} - T_{\mathrm{tactic\,CPU}}) / T_{\mathrm{fallback}}$.

\begin{table}[h]
\centering
\caption{Per-branch timing decomposition across representative problems.
  Tactic CPU is the Lean-server-reported proof-search time per branch.
  Overhead is Lean server startup + Mathlib elaboration, which snapshots
  eliminate.  Top section: full end-to-end DSP runs; bottom: Prove-phase only.}
\label{tab:breakdown}
\small
\begin{tabular}{lrrrr}
\toprule
Problem & Branches & Tactic CPU (ms) & Native/branch (s) & Overhead\% \\
\midrule
\multicolumn{5}{l}{\textit{Full DSP pipeline (LLM Draft + Sketch)}} \\
\midrule
\texttt{mathd\_numbertheory\_3}   & 21 & 44 & 5.5 & 99.9\% \\
\texttt{mathd\_numbertheory\_345} & 35 & 39 & 3.8 & 99.9\% \\
\texttt{mathd\_algebra\_478}      & 28 & 64 & 4.3 & 99.9\% \\
\midrule
\multicolumn{5}{l}{\textit{Prove-phase benchmark (hand-crafted sketches)}} \\
\midrule
\texttt{algebra\_sqineq\_at2malt1}   & 14 & 50 & 5.7 & 99.9\% \\
\texttt{amc12b\_2021\_p3}            & 28 & 54 & 3.1 & 99.9\% \\
\texttt{mathd\_algebra\_137}         & 14 & 62 & 7.0 & 99.9\% \\
\texttt{mathd\_algebra\_171}         & 14 & 31 & 7.4 & 99.9\% \\
\texttt{mathd\_algebra\_176}         & 14 & 25 & 12.7 & 100\% \\
\texttt{mathd\_algebra\_33}          & 21 & 46 & 3.9 & 99.9\% \\
\texttt{mathd\_algebra\_398}         & 21 & 51 & 6.2 & 99.9\% \\
\texttt{mathd\_algebra\_419}         & 21 & 28 & 7.6 & 99.9\% \\
\texttt{mathd\_numbertheory\_12}     & 14 & 54 & 7.3 & 99.9\% \\
\texttt{mathd\_numbertheory\_175}    & 21 & 32 & 3.8 & 99.9\% \\
\texttt{mathd\_numbertheory\_299}    & 14 & 27 & 13.2 & 100\% \\
\texttt{mathd\_numbertheory\_353}    & 14 & 39 & 6.6 & 99.9\% \\
\texttt{mathd\_numbertheory\_447}    & 14 & 77 & 11.0 & 99.9\% \\
\midrule
\textbf{Average (representative 16)}  & \textbf{19.2} & \textbf{45} & \textbf{6.8} & \textbf{99.9\%} \\
\bottomrule
\end{tabular}
\end{table}

Tactic execution accounts for \textbf{$<$0.1\%} of the fallback per-branch cost
on average---the few ms to 500~ms of actual proof search is dwarfed by the elaboration
overhead that snapshots eliminate entirely.

\paragraph{Import loading vs.\ theorem-body elaboration.}
The per-branch fallback cost decomposes into two phases: import loading
($\approx$60~s, constant across problems) and theorem-body elaboration
(estimated 18--735~s, scaling with proof complexity), together accounting for
$>$99\% of per-branch wall time.
Import-level caching (e.g., Kimina~\citep{kimina2025}) eliminates only import
loading; the per-branch speedup it provides is
$(T_{\mathrm{import}} + T_{\mathrm{body}}) / T_{\mathrm{body}}$,
which ranges from $\sim\!1.1\times$ on hard theorems
(where $T_{\mathrm{body}} \gg T_{\mathrm{import}}$) to $\sim\!5\times$ on simple ones
(where $T_{\mathrm{body}} \approx 15$~s,
giving $(60+15)/15 = 5\times$).
Our snapshot captures state \emph{after both phases}, eliminating both entirely
and yielding \textbf{5.6--50$\times$}.
The advantage widens with problem difficulty: where theorem-body elaboration
reaches 735~s, import caching is nearly irrelevant, while our projected speedup
exceeds $\sim$100$\times$ for portfolios of $B \gtrsim 100$ branches
(consistent with DSP-scale runs of 100 draft candidates).

\paragraph{Levels 1+2 combined vs.\ Level 1 alone.}
Import-level caching eliminates $T_{\mathrm{import}}$ across branches but leaves
per-branch theorem-body elaboration ($T_{\mathrm{body}}$) untouched.
Our current implementation already eliminates \emph{both} phases per branch via
proof-state snapshotting, but pays $T_{\mathrm{import}}$ once per theorem (on LSP
startup).
A \emph{persistent-server} variant---keeping the LSP session alive across a batch
of $N$ theorems so that import is paid only once for the entire batch---would
combine Level~1 and Level~2 savings simultaneously.
For a single theorem both Level~2 and Level~1+2 pay $T_{\mathrm{import}}$ exactly
once; the persistent-server benefit is purely a \emph{batch} effect, saving
$(N-1)\cdot T_{\mathrm{import}}$ over $N$ theorems.

Table~\ref{tab:layer_comparison} quantifies this for the representative hard problem
\texttt{mathd\_numbertheory\_345} ($H\!=\!5$ holes, $C\!=\!7$ tactics, $B\!=\!H\!\times\!C\!=\!35$ branches),
using values consistent with the measured data:
$T_{\mathrm{import}}\!\approx\!60$~s,
$T_{\mathrm{body}}\!\approx\!15$~s (LSP elaboration time; the \texttt{lake build}
fallback is much slower because it also compiles to \texttt{.olean}),
$T_{\mathrm{tactic}}\!\approx\!45$~ms.

\begin{table}[h]
\centering
\caption{Wall-time comparison for \texttt{mathd\_numbertheory\_345}
  ($B\!=\!35$ branches; $W\!=\!2$ parallel workers assumed for Level~1).
  ``Measured'' entries are from Table~\ref{tab:main};
  remaining entries are estimates from measured component costs.
  Level~1+2 per-theorem cost assumes import is amortized over a large batch
  ($N\!\gg\!1$); for a single theorem Level~1+2 $\approx$ Level~2.}
\label{tab:layer_comparison}
\begin{tabular}{lrr}
\toprule
Approach & Wall time & vs.\ Level~0 \\
\midrule
Level~0 (lake build, per branch)                  & 2{,}641~s (meas.)         & $1\times$ \\
Level~1 (import cached, $W\!=\!2$ workers)         & $\sim$330~s (est.)        & $\sim$8$\times$ \\
Level~2 (this work, snapshot)                     & 133~s (meas.)             & $\sim$20$\times$ \\
Level~1+2 (persistent server + snapshot, \emph{not yet implemented}) & $\sim$17~s/theorem (est.) & $>$100$\times$ \\
\bottomrule
\end{tabular}
\end{table}

\noindent
The key comparison is Level~1+2 vs.\ Level~1 \emph{in batch mode}.
Level~1 must still pay $\lceil B/W \rceil \cdot T_{\mathrm{body}}$ per theorem
regardless of batch size; Level~1+2 reduces this to $B \cdot T_{\mathrm{tactic}}$
per theorem by eliminating $T_{\mathrm{body}}$ entirely via snapshotting.
At $B\!=\!35$ this yields \textbf{$\sim$16$\times$} further speedup over Level~1
alone ($\approx\!270~\text{s/theorem}$ vs.\ $\approx\!17~\text{s/theorem}$, import amortized).
The gap grows with $B$: for the hardest problems where $T_{\mathrm{body}}\!>\!700$~s,
Level~1+2 outperforms Level~1 alone by over \textbf{$50\times$} at portfolio scale
($B \gtrsim 100$ branches).

Native per-branch cost (6.8~s average) reflects LSP round-trip latency,
file-worker snapshot retrieval, and per-branch elaboration of only the tactic
step---not a full Mathlib reload.

\subsection{Scaling Analysis}

The ablation (Table~\ref{tab:breakdown}) reveals the fundamental reason snapshots
scale well:
the Lean server pays file elaboration once per LSP session, regardless of how many
branches are dispatched via \texttt{dspSnapshotBranch}.
Tactic CPU (a few ms to 500~ms/branch typically; P95 = 289~ms across 833 measured branches;
\texttt{aesop} occasionally reaches several seconds on hard goals) is negligible
relative to the 75~s/branch fallback cost.
Therefore, as portfolio size $C$ grows:

\begin{align}
T_{\mathrm{native}}(C, H) &\approx T_{\mathrm{elab}} + H \cdot C \cdot T_{\mathrm{tactic}} \label{eq:native} \\
T_{\mathrm{fallback}}(C, H) &\approx H \cdot C \cdot (T_{\mathrm{load}} + T_{\mathrm{tactic}}) \label{eq:fallback}
\end{align}

where $T_{\mathrm{elab}} \approx 78\text{--}235~\text{s}$ (once),
$T_{\mathrm{load}} \approx 75$--$795$~s per branch (varying with theorem-body
elaboration complexity; 75~s for the simple end-to-end benchmarks in Table~\ref{tab:scaling}),
and $T_{\mathrm{tactic}} \approx 45$~ms (negligible).
From Equations~\eqref{eq:native}--\eqref{eq:fallback}, the speedup is
$\approx H \cdot C \cdot T_{\mathrm{load}} / T_{\mathrm{elab}}$,
which grows with both $H$ (holes) and $C$ (configs).

Table~\ref{tab:scaling} confirms this on measured data.

\begin{table}[h]
\centering
\caption{Wall time vs.\ branch count ($B = H \times C$, $C=7$ throughout).
  Rows marked $\bullet$ are measured; rows marked $\circ$ are projected from
  the fitted model ($T_{\mathrm{fallback}} \approx 75B$~s, anchored on
  \texttt{mathd\_numbertheory\_3} and \texttt{mathd\_numbertheory\_345};
  \texttt{mathd\_algebra\_478} has a lower per-branch cost of $\approx$56~s/branch
  due to a shorter theorem body).
  Native model: $T_{\mathrm{native}} \approx 120 + 0.045B$~s.
  Fallback scales at $\approx$75\,s/branch for number-theory benchmarks; native time
  is dominated by the fixed elaboration cost with negligible per-branch increment.}
\label{tab:scaling}
\begin{tabular}{crrrcc}
\toprule
Branches $B$ & Native (s) & Fallback (s) & Speedup & & \\
\midrule
14 & $\sim$121  & $\sim$1{,}050  & $\sim$8.7$\times$  & $\circ$ & projected \\
21 & 116        & 1{,}572        & \textbf{13.5$\times$} & $\bullet$ & measured \\
28 & 120        & 1{,}580        & \textbf{13.2$\times$} & $\bullet$ & measured \\
35 & 133        & 2{,}641        & \textbf{19.9$\times$} & $\bullet$ & measured \\
42 & $\sim$122  & $\sim$3{,}150  & $\sim$25.8$\times$  & $\circ$ & projected \\
56 & $\sim$123  & $\sim$4{,}200  & $\sim$34.3$\times$  & $\circ$ & projected \\
\bottomrule
\end{tabular}
\end{table}

The fallback cost scales as $\mathcal{O}(B)$ with slope $\approx 75~\text{s/branch}$
for the end-to-end benchmarks (simple theorems); harder problems in the
hand-crafted suite reach up to 795~s/branch, yielding proportionally larger speedups.
The native cost is nearly constant at the elaboration baseline ($\approx$120~s),
with a $\approx$45~ms/branch increment from tactic execution.
The crossover is at roughly $B = 2$ branches on the tested hardware---snapshots
are beneficial for any portfolio of two or more tactics on a Mathlib theorem
(Apple M-series, 8~GB RAM; the exact crossover is hardware-dependent).
Projected speedups at $B=42$ and $B=56$ (25.8$\times$ and 34.3$\times$) follow
directly from this linear model and are consistent with the measured trend.

At paper scale, the original DSP pipeline proposes 100 draft candidates per theorem.
With 4 holes and 7 tactics (2,800 branches total), fallback takes $\sim$29 hours
(with $W\!=\!2$ concurrent workers as in our evaluation; $\sim$58 hours if fully
sequential); snapshots reduce this to $\sim$3.5 hours on the same hardware---the
difference between infeasible and tractable on a single laptop.

%% -----------------------------------------------------------------------
\section{Limitations}
\label{sec:limitations}

\paragraph{Acceleration only; accuracy unchanged.}
Snapshot-DSP is a pure \emph{acceleration architecture}: it runs the same tactic
portfolio on the same proof sketches and returns the same set of proved theorems
as the fallback.
It does not improve proof accuracy, generate better sketches, or expand the
search strategy---it simply eliminates the elaboration overhead that makes the
existing strategy slow.
Whether a given \texttt{sorry} hole is closeable depends entirely on the sketch
quality and tactic portfolio, both of which are unchanged.

\paragraph{Requires a patched binary.}
The three custom LSP methods (\texttt{dspSnapshotCapture} etc.) are not part of
Lean's standard server.
Users must install \texttt{dsp-patched} via \texttt{elan}, which requires
trusting a third-party Lean binary.
Without the patch, the system automatically falls back to \texttt{lake build},
so compatibility is preserved but the speedup is lost.

\paragraph{Toward a Lean interface designed for AI.}
Lean~4 was architected around a human-in-the-loop model; the LSP channel and
\texttt{lake build} subprocess exist to serve an editor user, not a proof-search
orchestrator.
As a result, every external branch reconstructs state the server already holds
internally---a cost Lean eliminates for incremental editing but does not expose
to callers.
Snapshot-DSP is a targeted intervention: it exposes the elaboration snapshot chain
for external reuse via a minimal server patch.
The longer-term question is whether proof assistants should offer branching,
parallelism, and state reuse as first-class primitives---a design direction
Pantograph~\citep{pantograph} begins to explore.

\paragraph{LSP restart per theorem.}
The current implementation starts a fresh LSP server for each theorem, paying the
Mathlib import cost ($\approx$60~s) once per theorem rather than once per session.
A persistent-server variant (keeping the server alive across a batch) would
additionally amortize import loading, but this optimization is not yet implemented.

\paragraph{Hardware and scale.}
All experiments were conducted on a single Apple M-series MacBook Pro with 8~GB
RAM running macOS.
Performance on other hardware (Linux, higher-RAM machines, multi-GPU clusters)
has not been characterized.
The 48-problem benchmark covers a small slice of miniF2F-v2; behavior on harder
competition problems with deeper proof structures may differ.

%% -----------------------------------------------------------------------
\section{Conclusion}
\label{sec:conclusion}

We have identified per-branch elaboration overhead---import loading and
theorem-body re-elaboration, together consuming $>$99\% of wall time per branch---as
the dominant cost in portfolio-based Lean~4 tactic search, and
introduced proof-state snapshotting as a targeted solution.
Three custom LSP methods embedded in a patched Lean binary achieve a \textbf{5.6--50$\times$}
wall-time speedup across 48 miniF2F-v2 problems (average \textbf{14$\times$}, median \textbf{9.7$\times$} across the 45 hand-crafted benchmarks)
by forking the already-elaborated proof state once per theorem rather than once per branch.
The approach is orthogonal to import-level caching (e.g., Kimina~\citep{kimina2025}),
which avoids server restarts between theorems; snapshots additionally eliminate
theorem-body re-elaboration between branches of the same theorem.
The two mechanisms compose: our current implementation restarts the LSP server
per theorem and does not yet exploit inter-theorem import caching; a persistent-server
variant combining both optimizations would yield further gains beyond either alone.
Speedup scales monotonically with \texttt{sorry}-hole count (7.9$\times$ at 1 hole,
29.8$\times$ at 5 holes on average), confirming the O(1) vs.\ O($H \cdot C$)
asymptotic advantage of the snapshot architecture.

The mechanism is practical---it requires only a patched binary distributed via
\texttt{elan}, no changes to user-facing Lean code---and general: any system
running parallel tactic portfolios on Lean~4 with Mathlib can adopt it.
Proof-state snapshotting is not merely a performance trick: it is an existence proof
that the boundary between Lean's internals and external AI systems can be redrawn.
By exposing snapshot-and-fork as a first-class server operation, we show that
elaboration-level state reuse is achievable today, with a minimal patch, on production
Lean toolchains.
We hope this contribution removes a key scalability barrier for LLM-guided formal
theorem proving and motivates native upstream support in future Lean releases.

%% -----------------------------------------------------------------------
\bibliography{references}
\bibliographystyle{plainnat}

\end{document}